\documentclass[fleqn,twoside]{article}
\usepackage{amsmath,espcrc2,graphicx}

\def\pt{\widetilde p}
\def\pa{p_A}
\def\pn{p_N}

\begin{document}

\pagestyle{empty}

\begin{flushleft}
\Large{SAGA-HE-196-03     \hfill June 30, 2003}
\end{flushleft}
\vspace{2.4cm}
 
\begin{center}
\LARGE{\bf Nuclear modification of \\ 
           structure functions in lepton scattering} \\
\vspace{1.5cm}
\Large{ S. Kumano $^*$}  \\
\vspace{0.4cm}
{Department of Physics \\
 Saga University \\
 Saga, 840-8502, Japan} \\
\vspace{2.5cm}
\Large{Talk at the Second International Workshop on \\
       Neutrino-Nucleus Interactions in the Few GeV Region}

\vspace{0.6cm}
{UC Irvine, USA, December 12 - 15, 2002}

\vspace{0.5cm}
{(talk on Dec. 13, 2002) }  \\
\end{center}
\vspace{2.4cm}
\noindent{\rule{6.0cm}{0.1mm}} \\
\vspace{-0.3cm}
\normalsize

\noindent
{* Email: kumanos@cc.saga-u.ac.jp. URL: http://hs.phys.saga-u.ac.jp.}  \\

\vspace{+0.5cm}
\hfill {\large to be published in Nucl. Phys. B Supplements}
\vfill\eject
\setcounter{page}{1}
\pagestyle{plain}


\title{Nuclear modification of structure functions in lepton scattering}
\author{S. Kumano
\address{Department of Physics, Saga University,
              Saga, 840-8502, Japan}
\thanks{kumanos@cc.saga-u.ac.jp, http://hs.phys.saga-u.ac.jp}}
\begin{abstract}
We discuss nuclear structure functions in lepton scattering including
neutrino reactions. First, the determination of nuclear parton distribution
functions is explained by using the data of electron and muon deep
inelastic scattering and those of Drell-Yan processes. 
Second, NuTeV $sin^2 \theta_W$ anomaly is discussed by focusing on
nuclear corrections in the iron target.
Third, we show that the HERMES effect, which indicates nuclear modification
of the longitudinal-transverse structure function ratio, should exist
at large $x$ with small $Q^2$ in spite of recent experimental denials
at small $x$.
\vspace{1pc}
\end{abstract}

\maketitle
\section{Introduction}

Modification of nuclear structure functions or nuclear parton distribution
functions (NPDFs) is known especially in electron and muon scattering.
In neutrino scattering, such nuclear effects have not been seriously
investigated due to the lack of accurate deuteron data. Nuclear effects
in the PDFs have been investigated mainly among hadron structure physicists.
However, demands for accurate NPDFs have been growing from other fields
in the recent years. In fact, one of the major purposes of this workshop
\cite{nuint02} is to describe neutrino-nucleus cross sections 
for long baseline neutrino experiments, so that neutrino oscillations
could be understood accurately \cite{sakuda,py}. 

In the near future, neutrino cross sections should be understood
within a few percent level for the oscillation studies \cite{sakuda,py}.
Because typical nuclear corrections in the oxygen nucleus are larger
than this level, they should be precisely calculated. In low-energy
scattering, the nuclear medium effects are discussed in connection with
nuclear binding, Fermi motion, short-range correlations, Pauli exclusion
effects, and other nuclear phenomena. In this paper, the nuclear corrections
are discussed in the structure functions and the PDFs by focusing on
the high-energy region. These studies are important
not only for the neutrino studies but also for other applications.
For example, they are used in heavy-ion physics \cite{heavy}
for understanding accurate initial conditions of heavy nuclei, so that
one could make a definitive statement, for example on quark-gluon plasma
formation, in the final state. They could be also used in understanding
nuclear shadowing mechanisms \cite{recent-shadow}.

In this paper, recent studies are explained on the nuclear effects
which are relevant to high-energy neutrino scattering. 
First, a recent NPDF $\chi^2$ analysis is reported. Although
the unpolarized PDFs in the nucleon have been investigated extensively
\cite{pdf}, the NPDFs are not well studied.
However, there are some studies to obtain optimum NPDFs by using
a simple parametrization form and nuclear scattering data
\cite{ekrs,hkm}. We explain the current situation.
Second, NuTeV $sin^2 \theta_W$ anomaly \cite{nutev02} is investigated 
in a conservative way, namely in terms of nuclear corrections
\cite{nutevmod,nucl-sinth,sk02,kulagin}.
The NuTeV collaboration obtained anomalously large $sin^2 \theta_W$.
Before discussing any new physics mechanisms \cite{new}, we should
exclude possible nuclear physics explanations. In particular, the used
target is the iron and it may cause complicated nuclear medium effects.
Third, the HERMES effect \cite{hermes00}, which is nuclear
modification of the longitudinal-transverse structure function ratio,
is investigated in a simple convolution model.
It is intended to show that such an effect should exist in the medium
and large $x$ regions \cite{ek03} in spite of recent experimental
denials at small $x$ \cite{ccfr01,hermes02}. In particular,
the nucleon Fermi motion in a nucleus could play an important role
for the nuclear modification.

This paper consists of the following.
In section \ref{npdf}, global NPDF analysis results are shown 
The $sin^2 \theta_W$ anomaly topic is discussed in section \ref{sin2th}.
The HERMES effect is explained in section \ref{hermes}.
The results are summarized in section \ref{sum}.

\section{Nuclear parton distribution functions}
\label{npdf}

The determination of the NPDFs is 
not still satisfactory in comparison with the one for the nucleon.
It is partly because enough data are not obtained for fixing each
distribution from small $x$ to large $x$.
For example, various scaling violation data are not available, unlike
the HERA data for the proton, at very small $x$ for fixing gluon
distributions. However, the determination of nuclear PDFs has been
awaited for describing high-energy nuclear scattering phenomena, including
neutrino-nucleus and heavy-ion reactions. 
Some efforts have been made to provide practical parametrizations
for the NPDFs, such as the ones by Eskola, Kolhinen, Ruuskanen,
Salgado \cite{ekrs} and the ones by the HKM analysis \cite{hkm}.
In the following, the NPDFs are discussed based on the latter study
in Ref. \cite{hkm}.

First, the parametrization form should be selected. From the studies 
of nuclear $F_2$ structure function ratios $F_2^A/F_2^D$, one knows
the existence of shadowing phenomena at small $x$, anti-shadowing
at $x\approx 0.2$, depletion at medium $x$,
and then a positive nuclear modification at large $x$.
In order to express such $x$ dependence, the following
functions are used for the initial NPDFs at $Q_0^2$=1 GeV$^2$:
\begin{align}
& 
f_i^A (x, Q_0^2) = w_i(x,A,Z) \, f_i (x, Q_0^2),
\nonumber \\
& 
w_i(x,A,Z)  = 1 + \left( 1 - \frac{1}{A^{1/3}} \right) 
\nonumber \\
& \ \ \ \ \ \ \ \ \ \ \ 
\times 
\frac{a_i(A,Z) +b_i x+c_i x^2 +d_i x^3}{(1-x)^{\beta_i}}  .
\label{eqn:w}
\end{align}
Here, $Z$ is the atomic number, $A$ is the mass number, and
the subscript $i$ indicates a distribution type: $i$=$u_v$, $d_v$,
$\bar q$, or $g$. 
The functions $f_i^A$ and $f_i$ are the PDFs in a nucleus and the nucleon,
respectively, so that the weight function $w_i$ indicates
nuclear medium effects. The nuclear modification $w_i -1$ is assumed
to be proportional to $1-1/A^{1/3}$, and its $x$ dependence is taken
to be a cubic functional form with the $1/(1-x)^{\beta_i}$ factor
for describing the Fermi-motion part.
The parameters $a_i$, $b_i$, $c_i$, and $d_i$ are determined by
a $\chi^2$ analysis of experimental data.

Although the flavor dependence of the antiquark distributions
is known in the nucleon \cite{flavor}, the details of nuclear
antiquark distributions cannot be investigated at this stage.
Therefore, flavor symmetric antiquark distributions are assumed
in the parametrization.

The electron and muon deep inelastic experimental data and Drell-Yan
data are fitted by the NPDFs in Eq. (\ref{eqn:w}). The initial NPDFs
are, of course, evolved to various experimental $Q^2$ points,
and $\chi^2$ values are calculated in comparison with the data
for electron and muon deep inelastic scattering and Drell-Yan
processes:
\begin{equation}
\chi^2 = \sum_j \frac{(R_j^{data}-R_j^{theo})^2}
                     {(\sigma_j^{data})^2}.
\label{eqn:chi2}
\end{equation}
Here, $R$ is the ratio $F_2^A/F_2^{A'}$ or
$\sigma_{DY}^A/\sigma_{DY}^{A'}$.
These structure functions and the DY cross sections are
calculated in the leading order. The experimental error is
given by systematic and statistical errors as
$(\sigma_j^{data})^2 = (\sigma_j^{sys})^2 + (\sigma_j^{stat})^2$.
The first version was published in 2001, and then
the research is in progress by including the Drell-Yan data.
We discuss the obtained NPDFs by these analyses.

\begin{figure}[t!]
\begin{center}
     \includegraphics[width=0.45\textwidth]{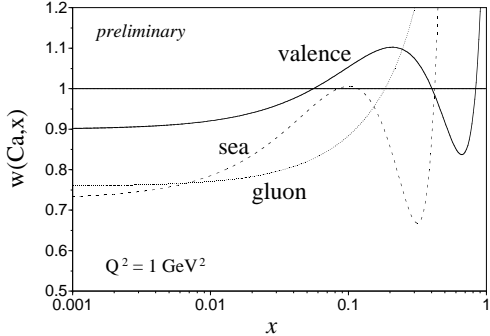}
\end{center}
\vspace{-1.3cm}
\caption{Obtained weight functions for the calcium nucleus at
         $Q^2$=1 GeV$^2$.}
\vspace{-0.1cm}
\label{fig:wxca1}
\end{figure}

Obtained optimum distributions are shown for the calcium nucleus
at $Q^2$=1 GeV$^2$ in Fig. \ref{fig:wxca1}.
The solid, dashed, and dotted curves indicate
the weight functions for the valence-quark, antiquark, and gluon
distributions. The valence distribution is well determined in
the medium $x$ region, but it is difficult to determine it at
small $x$ although it is constrained by the baryon-number and charge
conservations. In fact, it will be one of the NuMI projects \cite{numi}
to determine the valence-quark ($F_3$) shadowing in comparison with
the antiquark ($F_2$) shadowing by neutrino-nucleus scattering. 
On the other hand, the antiquark distribution is well determined
at small $x$; however, it cannot be fixed at medium $x$ ($x>0.2$) in spite
of the momentum-conservation constraint. Because this is the leading
order analysis, the gluon distribution is not fixed in the whole
$x$ region. 

The obtained NPDFs are available at the web site
http://hs.phys.saga-u.ac.jp/nuclp.html,
where computer codes are available for calculating the distributions at
given $x$ and $Q^2$ for a requested nucleus. The nuclear type should be
in the range, $2 \le A \le 208$, in principle, because the analyzed
nuclei are in this range. However, the distributions could be also calculated
for larger nuclei ($A>208$) because variations of the NPDFs are rather small
in such a large-$A$ region. If one wishes to use an analytical form,
the distributions at $Q^2$=1 GeV$^2$ are provided in the appendix of
Ref. \cite{hkm}. After the first version was published, a new analysis
has been investigated. The second version will become available within
the year of 2003.

\section{A nuclear physicist's view of $sin^2 \theta_W$ anomaly}
\label{sin2th}

The NuTeV collaboration announced that their measurement of
the weak mixing angle $sin^2 \theta_W$ is significantly different from
collider measurements. If the neutrino-nucleus scattering data
are excluded, a global analysis indicates 
$sin^2 \theta_W^{on-shell}= 0.2227 \pm 0.0004$ \cite{lep01}.
On the other hand, the NuTeV reported \cite{nutev02}
\begin{equation}
sin^2 \theta_W = 0.2277 \pm 0.0013 \, \text{(stat)} 
                         \pm 0.0009 \, \text{(syst)}
\, ,
\end{equation}
by using their neutrino and antineutrino scattering data.

Because it is one of the important constants in the standard model,
we should find a reason for the discrepancy. Although it may indicate the
existence of a new mechanism \cite{new}, we should seek a conservative
explanation first. In particular, the NuTeV target is the iron nucleus so
that nuclear medium effects might have altered the $sin^2 \theta_W$ value
\cite{nutevmod,nucl-sinth,sk02,kulagin}.
In the following, we explain nuclear effects on the $sin^2 \theta_W$
determination. 

The neutrino and antineutrino cross section data are analyzed
by a special Monte Carlo code, so that it is not theoretically
straightforward to investigate a possible explanation. 
In order to simplify the investigation, we study nuclear effects 
on the Paschos-Wolfenstein (PW) relation, which is considered to be
``implicitly" used in the NuTeV analysis. 
The PW relation \cite{pw} is given by the ratio of charged current (CC)
and neutral current (NC) cross sections:
\begin{equation}
R^-  = \frac{  \sigma_{NC}^{\nu N}  - \sigma_{NC}^{\bar\nu N} }
              {   \sigma_{CC}^{\nu N}  - \sigma_{CC}^{\bar\nu N} }
        =  \frac{1}{2} - sin^2 \theta_W 
\, .
\label{eqn:pw}
\end{equation}
This relation is valid for the isoscalar nucleon; however, 
corrections should be carefully investigated for the non-isoscalar
iron target.

If the relation is calculated for a nucleus in the leading order
of $\alpha_s$, we obtain \cite{sk02}
\begin{align}
R_A^- & = \frac{   \sigma_{NC}^{\nu A} - \sigma_{NC}^{\bar\nu A} }
                  {   \sigma_{CC}^{\nu A} - \sigma_{CC}^{\bar\nu A} }
\nonumber \\
       & \! \! \! \! \! \! \! \! \! \! 
 = \{ 1-(1-y)^2 \} \,  [ \, (u_L^2 -u_R^2 ) \{ u_v^A(x) + c_v^A (x) \}
\nonumber \\
       &  \ \ \ \ \ \ \ \ \ \ \ \ \ \ \ \ 
+  (d_L^2 -d_R^2 ) \{ d_v^A (x) + s_v^A (x) \} \, ]
\nonumber \\
       &  \! \! \! \! \! \! \! \!
 / \, [ \, d_v^A (x) + s_v^A (x) 
         - (1-y)^2 \, \{ u_v^A (x) + c_v^A (x) \} \, ]
\, ,
\label{eqn:apw1}
\end{align}
where the valence quark distributions are defined by
$q_v^A \equiv q^A -\bar q^A$. 
The couplings are expressed by the weak mixing angle as
$u_L = 1/2- (2/3) \, sin^2 \theta_W$,
$u_R = -(2/3) \, sin^2 \theta_W$,
$d_L = -1/2 +(1/3) \, sin^2 \theta_W$, and
$d_R = (1/3) \, sin^2 \theta_W$.
It is known that the nuclear distributions are modified from those
for the nucleon. The modification for $u_v^A$ and $d_v^A$ could be
expressed by the weight functions $w_{u_v}$ and $w_{d_v}$ at any $Q^2$:
\begin{align}
u_v^A (x) & = w_{u_v} (x,A,Z) \, \frac{Z \, u_v (x) + N \, d_v (x)}{A},
\nonumber \\
d_v^A (x) & = w_{d_v} (x,A,Z) \, \frac{Z \, d_v (x) + N \, u_v (x)}{A},
\label{eqn:wpart}
\end{align}
although $w_i$ in section 2 is defined at fixed $Q^2$ (=$Q_0^2$).
Here, $u_v$ and $d_v$ are the distributions in the proton,
and  $N$ is the neutron number.

In order to find possible deviation from the PW relation, we 
first define a function which is related to the neutron excess in
a nucleus:
$\varepsilon_n (x) = [(N-Z)/A] (u_v-d_v)/(u_v+d_v) $,
and then a difference between the weight functions is defined by
\begin{equation}
\varepsilon_v (x) = \frac{w_{d_v}(x,A,Z)-w_{u_v}(x,A,Z)}
                             {w_{d_v}(x,A,Z)+w_{u_v}(x,A,Z)}
\, .
\label{eqn:en}
\end{equation}
Furthermore, there are correction factors associated with
the strange and charm quark distributions, so that we define
$\varepsilon_s$ and $\varepsilon_c$ by
$\varepsilon_s = s_v^A /[w_v \, (u_v+d_v)]$ and
$\varepsilon_c = c_v^A /[w_v \, (u_v+d_v)]$
with $w_v = (w_{d_v}+w_{u_v})/2$.

Neutron-excess effects are taken into account in the NuTeV analysis
as explained by McFarland {\it et al.} \cite{nutevmod},
and they are also investigated by Kulagin \cite{kulagin}.
The strange quark ($\varepsilon_s$) contribution is small according to
Zeller {\it et al.} \cite{sv}, and it increases the deviation.
Here, we investigate a different contribution from the $\varepsilon_v (x)$
term \cite{sk02}. 
Writing Eq. (\ref{eqn:apw1}) in terms of the factors, 
$\varepsilon_n$, $\varepsilon_v$, $\varepsilon_s$, and $\varepsilon_c$, 
and then expanding the expressions by these small factors, we obtain
\begin{align}
& 
 R_A^-   =  \frac{1}{2} - sin^2 \theta_W  
\nonumber \\
&  
- \varepsilon_v (x)  \bigg \{ \bigg ( \frac{1}{2} - sin^2 \theta_W \bigg ) 
               \frac{1+(1-y)^2}{1-(1-y)^2} - \frac{1}{3} sin^2 \theta_W 
\bigg \}
\nonumber \\
& 
+O(\varepsilon_v^2)+O(\varepsilon_n)+O(\varepsilon_s)+O(\varepsilon_c) 
\, .
\label{eqn:apw3}
\end{align}
Because only the $\varepsilon_v$ contribution is discussed in the following,
other terms are not explicitly written in the above equation.
This equation indicates that the observed $sin^2 \theta_W$ in
neutrino-nucleus scattering is effectively larger if the ratio is calculated
without the $\varepsilon_v$ correction.

The nuclear modification difference $\varepsilon_v(x)$ is not known at all
at this stage. We try to estimate it theoretically by using charge and
baryon-number conservations:
$Z  = \int dx \, A   \sum_q e_q  (q^A - \bar q^A)$ and
$A  = \int dx \, A   \sum_q (1/3) \, (q^A - \bar q^A)$.
These equations are expressed by the valence-quark distributions,
then they becomes
\begin{align}
& 
\int dx \, (u_v+d_v) \,  [ \, \Delta w_v 
         +  w_v \, \varepsilon_v (x) \, \varepsilon_n (x) \, ] = 0
\, ,
\label{eqn:b} \\
& 
\int dx \, (u_v+d_v) \,  [ \, \Delta w_v \, 
         \{ 1-3 \, \varepsilon_n(x) \} \,
\nonumber \\
& \ \ \ \ \ \ \ \ \ \ \ \ \ \ 
        - w_v \, \varepsilon_v (x) \, 
         \{ 3 - \varepsilon_n (x) \}  \, ] =0
\, ,
\label{eqn:c}
\end{align}
where $\Delta w_v$ is defined by $\Delta w_v=w_v -1$.
These equations suggest that there should exist a finite distribution
for $\varepsilon_v(x)$ due to the charge and baryon-number conservations.
However, there is no unique solution for these integral equations, so that
the following discussions become inevitably model dependent.

\begin{figure}[h!]
\begin{center}
     \includegraphics[width=0.45\textwidth]{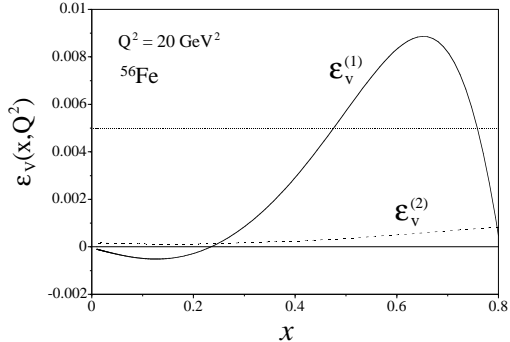}
\end{center}
\vspace{-1.3cm}
\caption{The function $\varepsilon_v(x,Q^2)$ is estimated by
         two different descriptions at $Q^2$=20 GeV$^2$.}
\vspace{-0.1cm}
\label{fig:epsv12}
\end{figure}

We provide two examples for estimating the order of magnitude of
the effect on $sin^2 \theta_W$. First, the integrands of
Eqs. (\ref{eqn:b}) and (\ref{eqn:c}) are assumed to vanish 
by neglecting the higher-order terms $O(\varepsilon_v \varepsilon_n)$:
\begin{equation}
\text{case 1:}\ \ 
\varepsilon_v (x) 
= - \varepsilon_n (x) \,  \frac{\Delta w_v(x)}{w_v(x)}
\, .
\label{eqn:evx2}
\end{equation}
Second, the $\chi^2$ analysis result \cite{hkm}, which is explained
in section \ref{npdf}, could be used 
for the estimation:
\begin{equation}
\text{case 2:}\ \ 
\varepsilon_v (x) = \left [ \frac{w_{d_v}(x)-w_{u_v}(x)}
                             {w_{d_v}(x)+w_{u_v}(x)} 
    \right ]_{\text{$\chi^2$ analysis}}
\, .
\label{eqn:ev}
\end{equation}
These two descriptions are numerically estimated and the results
are shown at $Q^2$=20 GeV$^2$ in Fig. \ref{fig:epsv12}.
The solid and dashes curves indicate the case 1 and 2, respectively.

In the first case, the function $\varepsilon_v$ is directly proportional
to the nuclear modification $\Delta w_v(x)$, so that it changes the sign
at $x \sim 0.2$. In comparison with the NuTeV deviation 0.005,
which is shown by the dotted line, 
$\varepsilon_v^{(1)}$ is of the same order of magnitude. 
On the other hand, the second one $\varepsilon_v^{(2)}$ is rather
small. This is partly because of the assumed functional form
in the $\chi^2$ analysis \cite{hkm}, which
was not intended especially to obtain the nuclear modification
difference $\varepsilon_v$. Because the distributions are much
different depending on the model, numerical estimates are merely 
considered to be an order of magnitude estimate.

\begin{figure}[h!]
\begin{center}
     \includegraphics[width=0.45\textwidth]{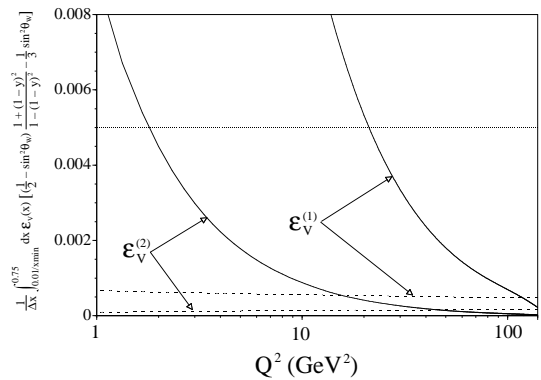}
\end{center}
\vspace{-1.3cm}
\caption{Contributions to $sin^2 \theta_W$ are calculated by
         taking the $x$ average and they are shown by the solid
         curves. The dashed curves are calculated by taking
         the NuTeV kinematics into account \cite{sk02}.}
\vspace{-0.1cm}
\label{fig:q2dep}
\end{figure}

As far as we see Fig. \ref{fig:epsv12}, our mechanism seems to a promising
explanation for the NuTeV anomaly. If a simple $x$ average is taken
for the $\varepsilon_v$ contribution to the $sin^2 \theta_W$ determination,
we obtain the solid curves in Fig. \ref{fig:q2dep}, and they are of
the order of the NuTeV deviation. However, the situation is not so simple. 
Although the function $\varepsilon_v^{(1)}$ is large in the large $x$ region
in Fig. \ref{fig:epsv12}, few NuTeV data exist in such a region. It means
that the $\varepsilon_v^{(1)}$ contribution to $sin^2 \theta_W$ could be
significantly reduced if the NuTeV kinematics is taken into account.
The guideline of incorporating such experimental
kinematics is supplied in Fig. 1 of Ref. \cite{sv}.
The distribution $\varepsilon_v$ could be effectively simulated
by the NuTeV functions, $u_v^p-d_v^p$ and  $d_v^p-u_v^p$, although
physics motivation is completely different. Using the NuTeV functionals
\cite{sv,mz}, we obtain the dashed curves in Fig. \ref{fig:q2dep}.
Because of the lack of large $x$ data, the contributions are significantly
reduced. 

The mechanism due to the nuclear modification difference 
between $u_v$ and $d_v$ could partially explain the NuTeV deviation,
but it is not a major mechanism for the deviation according to
Fig. \ref{fig:q2dep}. However, the distribution $\varepsilon_v$
itself is not known at all, so that it would be too early to exclude
the mechanism. On the other hand, it should be an interesting topic
to investigate $\varepsilon_v$ experimentally by the NuMI \cite{numi}
and neutrino-factory \cite{nufact} projects.

\section{HERMES effect}
\label{hermes}

The HERMES effect indicates nuclear modification of 
the longitudinal-transverse structure function ratio $R(x,Q^2)$.
It was originally reported at small $x$ ($0.01<x<0.03$) with small $Q^2$
($0.5<Q^2<1$ GeV$^2$) in the HERMES paper in 2000  \cite{hermes00}.
There are theoretical investigations on this topic in terms of
shadowing \cite{shadow} and an isoscalar meson \cite{mbk00}.

This interesting nuclear effect is, however, not observed in the CCFR/NuTeV
experiments \cite{ccfr01}. Although the CCFR/NuTeV target is the iron nucleus,
observed R values agree well with theoretical calculations for the nucleon
in the same kinematical region with the HERMES.
Furthermore, a more careful HERMES analysis of radiative corrections
showed such modification does not exist anymore \cite{hermes02}. 

Considering these experimental results, one may think that such a nuclear
effect does not exist at all. However, we point out that the effect should
exist in a different kinematical region, namely at large $x$ with small $Q^2$ 
\cite{ek03}. The existence of a nuclear effect in $R(x,Q^2)$ is important
not only for investigating nuclear structure in the parton model
but also for many analyses of lepton scattering data. For example,
the SLAC parametrization in 1990 \cite{r1990} has been used as a popular
one. However, the data contain nuclear ones, so that one cannot
use it for nucleon scattering studies if large nuclear effects exist
in the data. 
Because of the importance of $R(x,Q^2)$ in lepton scattering analyses,
we investigate the possibility of nuclear modification theoretically.

The structure function for the photon polarization $\lambda$ is
$W^{A,N}_\lambda = \varepsilon_\lambda^{\mu *} \,
                  \varepsilon_\lambda^\nu     \,  W^{A,N}_{\mu\nu}$,
so that longitudinal and transverse ones are defined by
$ W^{A,N}_T = ( W^{A,N}_{+1} + W^{A,N}_{-1} ) /2$ and
$ W^{A,N}_L = W^{A,N}_0 $. Here, $N$ and $A$ denote the nucleon and
a nucleus, respectively. Lepton-hadron scattering cross section is
described by a lepton tensor multiplied by a hadron tensor $W_{\mu \nu}$.
In the electron scattering, the tensors for the nucleon and
a nucleus are given by
\begin{align}
W^{A,N}_{\mu\nu} (p_{_{A,N}},  &  q)  =
  - W^{A,N}_1 (p_{_{A,N}}, q) 
  \left ( g_{\mu\nu} - \frac{q_\mu q_\nu}{q^2} \right )
\nonumber \\
& + W^{A,N}_2 (p_{_{A,N}}, q) \, \frac{\pt_{_{A,N} \mu} 
  \, \pt_{_{A,N} \nu}}{p_{_{A,N}}^2} 
\ ,
\label{eqn:hadron}
\end{align}
where $\pt_{\mu} = p_{\mu} -(p \cdot q) \, q_\mu /q^2$.
In terms of these structure functions, the longitudinal and transverse
structure functions are given 
$ W^{A,N}_T = W^{A,N}_1 $ and
$ W^{A,N}_L = (1+\nu_{_{A,N}}^2/Q^2) W^{A,N}_2 - W^{A,N}_1 $
by taking the nucleus or nucleon rest frame.
Here, $\nu_A \equiv \nu$, and
the photon momentum in the nucleon rest frame is 
denoted $(\nu_N,\vec q_N)$ with $\nu_N^2 = (p_N \cdot q)^2 /p_N^2$.

We use a conventional convolution description
for nuclear structure functions:
\begin{equation}
W^A_{\mu\nu} (\pa, q) = \int d^4 \pn \, S(\pn) \, W^N_{\mu\nu} (\pn, q)
\ ,
\label{eqn:conv}
\end{equation}
where $p_N$ is the nucleon momentum and $S(p_N)$ is the spectral
function which indicates the nucleon momentum distribution in a nucleus. 
In order to investigate the longitudinal and transverse components,
we introduce projection operators which
satisfy $\widehat P_1^{\, \mu\nu} W^A_{\mu\nu} = W_1^A$ and
$\widehat P_2^{\, \mu\nu} W^A_{\mu\nu} = W_2^A$. 
They are explicitly written as
$ \widehat P_1^{\, \mu\nu} = - (1/2)
   \left ( g^{\mu\nu} - \pt_A^{\, \mu} \, \pt_A^{\, \nu} 
                        / \pt_A^{\, 2} \right ) $ and
$ \widehat P_2^{\, \mu\nu} = - p_A^2 / (2\, \pt_A^{\, 2})
  \left ( g^{\mu\nu} - 3 \, \pt_A^{\, \mu} \, \pt_A^{\, \nu}
                        / \pt_A^{\, 2} \right ) $.
Instead of $W_1$ and $W_2$ structure functions, the functions
$F_1$ and $F_2$ are usually used:
$F_1^{A,N} = \sqrt{p_{_{A,N}}^2} \, W_1^{A,N}$ and
$F_2^{A,N} = ( p_{_{A,N}} \cdot q / \sqrt{p_{_{A,N}}^2}) \, W_2^{A,N} $.
Then, the longitudinal structure function is given by
\begin{align}
F_L^{A,N} (x_{_{A,N}}, Q^2) & = \bigg ( 1 + \frac{Q^2}{\nu_{_{A,N}}^2} \bigg ) 
        F_2^{A,N} (x_{_{A,N}}, Q^2)
\nonumber \\
&                    - 2 x_{_{A,N}} F_1^{A,N} (x_{_{A,N}}, Q^2)
\, ,
\label{eqn:flla}
\end{align}
where
$x_A = Q^2 /(2 \, p_A \cdot q)$ and
$x_N = Q^2 /(2 \, p_N \cdot q)$.
The ratio $R_A$ of the longitudinal cross section to the transverse one
is expressed by the function $R_A(x_A,Q^2)$:
\begin{equation}
R_A (x_A, Q^2) = \frac{F_L^A (x_A, Q^2)}{2 \, x_A F_1^A (x_A, Q^2)} 
\ .
\end{equation}

Applying the projection operators $\widehat P_1^{\, \mu\nu}$ and
$\widehat P_2^{\, \mu\nu}$ to Eq. (\ref{eqn:conv}), we have
\begin{align}
& \! \! \! 
2 \, x_A F_1^A (x_A, Q^2) =  \int d^4 \, p_N \, S(p_N) \, z \, 
\frac{M_N}{\sqrt{p_N^2}} 
\nonumber \\
& \! \! \! \! 
 \times
\bigg [ \bigg ( 1 
            + \frac{\vec p_{N\perp}^{\ 2}}{2 \, \pt_N^{\, 2}} \bigg )
             2  x_N F_1^N (x_N, Q^2)
+  \frac{\vec p_{N\perp}^{\ 2}}{2  \pt_N^{\, 2}}
             F_L^N (x_N, Q^2)  \bigg ]
\, ,
\label{eqn:trans}
\\
& \! \! \! 
F_L^A (x_A, Q^2) =  \int d^4 \, p_N  \, S(p_N) \, z \, 
\frac{M_N}{\sqrt{p_N^2}}
\nonumber \\
& \! \! \! \! 
\times
\bigg [ \bigg ( 1 
            + \frac{\vec p_{N\perp}^{\ 2}}{\pt_N^{\, 2}} \bigg )
             F_L^N (x_N, Q^2)
+  \frac{\vec p_{N\perp}^{\ 2}}{\pt_N^{\, 2}}
            2  x_N F_1^N (x_N, Q^2) \bigg ]
\, .
\label{eqn:longi}
\end{align}
These results are interesting. The transverse structure function
for a nucleus is described not only by the transverse one for
the nucleon but also by the longitudinal one 
with the admixture coefficient 
$\vec p_{N\perp}^{\ 2}/(2  \pt_N^{\, 2})$. 
The $\vec p_{N\perp}$ is the nucleon momentum 
component perpendicular to the photon one $\vec q$.
Equations (\ref{eqn:trans}) and  (\ref{eqn:longi}) indicate that
the transverse-longitudinal admixture exists because the nucleon
momentum direction is not necessary along the virtual photon direction.

\begin{figure}[h!]
\begin{center}
     \includegraphics[width=0.45\textwidth]{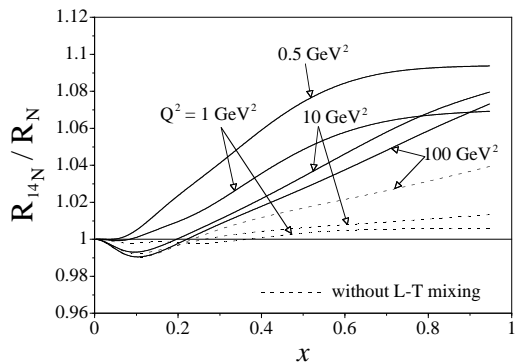}
\end{center}
\vspace{-1.3cm}
\caption{The nitrogen-nucleon ratio $R_{^{14}N}/R_N$ is shown
         at $Q^2$=0.5, 1, 10, and 100 GeV$^2$. The solid curves are the
         full results and the dashed ones are obtained by
         terminating the admixture effects.}
\vspace{-0.1cm}
\label{fig:rratio03}
\end{figure}

These expressions are numerically estimated for the nitrogen nucleus
by taking a simple shell model for the spectral function with
density dependent Hartree-Fock wave functions. Parton distribution
functions are taken from the MRST-1998 version and the nucleonic
$R(x,Q^2)$ is taken from the SLAC analysis in 1990 \cite{r1990}.
The nitrogen-nucleon ratios $R_{^{14}N}/R_N$ are shown at
$Q^2$=1, 10, 100 GeV$^2$ by the solid curves in Fig. \ref{fig:rratio03}.
In order to clarify the admixture effects, the ratios are also
calculated by suppressing the $\vec p_{N\perp}^{\ 2}$ terms, 
and the results are shown by the dashed curves.
In addition, the nuclear modification is calculated at $Q^2$=0.5 GeV$^2$
by using the GRV94 parametrization for the PDFs. It is intended to find
the modification magnitude at smaller $Q^2$, where 
JLab experiments could possibly probe \cite{jlab}.
Because the admixture is proportional to 
$\vec p_{N\perp}^{\ 2}/(2  \pt_N^{\, 2}) \sim \vec p_{N\perp}^{\ 2}/Q^2$,
the modification effects are large at small $Q^2$ (=0.5 $-$ 1 GeV$^2$)
and they become small at large $Q^2$. However, the modification
does not vanish even at $Q^2$=100 GeV$^2$ due to the Fermi-motion 
and binding effects which are contained implicitly in the spectral
function.

In this way, we found that the nucleon Fermi motion, especially
the perpendicular motion to the virtual photon direction, and 
the nuclear binding give rise to the nuclear modification of
the longitudinal-transverse ratio $R(x,Q^2)$.
However, nuclear modification of $R$ in the large $x$ region with
small $Q^2$ has not been investigated experimentally. The situation
is clearly illustrated in Fig. 3 of Ref. \cite{ccfr01}, where
the data does not exist at $x=0.5$ with $Q^2 \approx$1 GeV$^2$.
We hope that future measurements, for example those of JLab
experiments \cite{jlab}, are able to provide clear information on
the nuclear modification in this region.

\section{Summary}
\label{sum}

Current neutrino scattering experiments are done with nuclear targets,
so that precise nuclear corrections should be taken into account
in order to investigate underlying elementary processes, for example
neutrino oscillation phenomena. In this paper, the discussions are
focused on high-energy reactions. 

First, the optimum nuclear parton distribution functions were 
determined by the $\chi^2$ analysis of DIS and Drell-Yan
data. They could be used for calculating high-energy nuclear
cross sections.

Second, a possibility of explaining the NuTeV $sin^2 \theta_W$
was investigated by the nuclear correction difference between $u_v$ and $d_v$
in the iron nucleus. Although the contribution to the $sin^2 \theta_W$
deviation may not be large at this stage, the distribution 
$\varepsilon_v (x)$ should be investigated by future experiments.

Third, a possible HERMES-type effect was proposed
in the medium and large $x$ regions due to the nucleon Fermi motion
and binding. Especially, we found that the perpendicular nucleon motion
to the virtual photon direction gives rise to the admixture of longitudinal
and transverse structure functions in the nucleon. Such an effect should
be tested by electron and neutrino scattering experiments at large $x$
with small $Q^2$.

\section*{Acknowledgments}
S.K. was supported by the Grant-in-Aid for Scientific
Research from the Japanese Ministry of Education, Culture, Sports,
Science, and Technology. He thanks M. Sakuda for his financial support 
for participating in this workshop.



\begin{thebibliography}{9}
\bibitem{nuint02} http://nuint.ps.uci.edu.
\bibitem{sakuda} M. Sakuda, http://nuint.ps.uci.edu/slides/ Sakuda.pdf;
                 in proceedings of this workshop.
\bibitem{py} E. A. Paschos and J. Y. Yu, Phys. Rev. D65 (2002) 033002. 
\bibitem{heavy} Shi-yuan Li and Xin-Nian Wang, Phys. Lett. B527 (2002) 85;
        X. Zhang and G. Fai, Phys. Rev. C65 (2002) 064901;
        A. Chamblin and G. C. Nayak, Phys. Rev. D66 (2002) 091901.
 \bibitem{recent-shadow} 
        B. Z. Kopeliovich and A. V. Tarasov, Nucl. Phys. A710 (2002) 180;
        L. Frankfurt, V. Guzey, and M. Strikman, hep-ph/0303022;
        N. Armesto {\it et. al.}, hep-ph/0304119.
\bibitem{pdf}   http://durpdg.dur.ac.uk/hepdata/pdf.html.
\bibitem{ekrs}  K. J. Eskola, V. J. Kolhinen, and P. V. Ruuskanen,
                         Nucl. Phys. B535 (1998) 351;
                    K. J. Eskola, V. J. Kolhinen, and C. A. Salgado,
                         Eur. Phys. J. C9 (1999) 61. 
\bibitem{hkm} M. Hirai, S. Kumano, and M. Miyama,
                     Phys. Rev. D64 (2001) 034003; research in progress.
                 See http://hs.phys.saga-u.ac.jp/nuclp.html.
\bibitem{nutev02} G. P. Zeller {\it et. al.},
                   Phys. Rev. Lett. 88 (2002) 091802.
\bibitem{nutevmod} K. S. McFarland {\it et. al.}, 
                      Nucl. Phys. B 112 (2002) 226.
\bibitem{nucl-sinth} G. A. Miller and A. W. Thomas, hep-ex/0204007;
       G. P. Zeller {\it et. al.}, hep-ex/0207052.
       W. Melnitchouk and A. W. Thomas, Phys. Rev. C67 (2003) 038201;
       S. Kovalenko, I. Schmidt, and J.-J. Yang, Phys. Lett. B546 (2002) 68.
\bibitem{sk02} S. Kumano, Phys. Rev. D66 (2002) 111301.
\bibitem{kulagin} S. A. Kulagin, Phys. Rev. D67 (2003) 091301.
\bibitem{new}  S. Davidson {\it et. al.}, 
                   J. High Energy Phys. 0202, 037 (2002);
               E. Ma and D. P. Roy,
                   Phys. Rev. D65 (2002) 075021;
               C. Giunti and M. Laveder, hep-ph/0202152;
               W. Loinaz, N. Okamura, T. Takeuchi, and L. C. R. Wijewardhana,
                   Phys. Rev. D67 (2003) 073012.
\bibitem{hermes00} K. Ackerstaff {\it et al.}, Phys. Lett. B475  (2000) 386.
\bibitem{ek03}  M. Ericson and S. Kumano, Phys. Rev. C 67 (2003) 022201. 
\bibitem{ccfr01} U. K. Yang {\it et al.}, Phys. Rev. Lett. 87 (2001)  251802. 
\bibitem{hermes02} A. Airapetian {\it et al.}, hep-ex/0210067 \& 0210068.
\bibitem{flavor} S. Kumano, Phys. Rep. 303 (1998) 183;
                 G. T. Garvey and J.-C. Peng,
                       Prog. Part. Nucl. Phys. 47 (2001) 203.
\bibitem{numi} J. G. Morfin, Nucl. Phys. B112 (2002) 251.
\bibitem{lep01}  D. Abbaneo {\it et. al.}, hep-ex/0112021. 
                  See also the reference [21] in Ref. \cite{nutev02}.
\bibitem{pw} E. A. Paschos and L. Wolfenstein, 
                   Phys. Rev. D7 (1973) 91.
\bibitem{sv}      G. P. Zeller {\it et. al.}, 
                      Phys. Rev. D65 (2002) 111103 .
\bibitem{mz} K. S. McFarland and G. P. Zeller, personal communications.
\bibitem{nufact} http://www.cap.bnl.gov/nufact03/.
\bibitem{shadow}  V. Barone and M. Genovese, hep-ph/9610206;
                  B. Kopeliovich, J. Raufeisen, and A. Tarasov,
                     Phys. Rev. C62 (2000) 035204.
\bibitem{mbk00}    G. A. Miller, S. J. Brodsky, and M. Karliner,
                     Phys. Lett. B481 (2000) 245;
                   G. A. Miller, Phys. Rev. C64 (2001) 022201.
\bibitem{r1990} L. W. Whitlow, S. Rock, A. Bodek, S. Dasu,
                    and E. M. Riordan,
                    Phys. Lett. B250 (1990) 193;
                L. W. Whitlow, report SLAC-357 (1990).
\bibitem{jlab} H. P. Blok, personal communications.
               A. Br\"ull {\it et al.},
               http://www.jlab.org/exp\_prog /proposals/99/PR99-118.pdf.
\end{thebibliography}
\end{document}